\documentclass[prc,aps,preprint,superscriptaddress,showpacs]{revtex4}
\usepackage{epsfig}
\usepackage{amsmath}
\usepackage{amssymb}
\usepackage{bm}

\date{\today}
\newcommand{\be}{\begin{equation}}
\newcommand{\ee}{\end{equation}}
\newcommand{\beq}{\begin{eqnarray}}
\newcommand{\eeq}{\end{eqnarray}}
\newcommand{\bc}{\begin{center}}
\newcommand{\ec}{\end{center}}
\newcommand{\ba}{\begin{array}}
\newcommand{\ea}{\end{array}}
\begin{document}
\title{Cranking in isospace}

\author{{\sc Sebastian G{\l}owacz}}
\affiliation{Institute of Theoretical Physics, University of Warsaw,\\
             ul. Ho{\.z}a 69, PL-00 681 Warsaw, Poland}
\author{{\sc  Wojciech Satu{\l}a}}
\email{             satula@fuw.edu.pl}
\affiliation{Institute of Theoretical Physics, University of Warsaw,\\
             ul. Ho{\.z}a 69, PL-00 681 Warsaw, Poland}
\affiliation{          KTH (Royal Institute of Technology),\\
           AlbaNova University Centre,106 91 Stockholm}
\affiliation{          Department of Physics, University of Tennessee,
              Knoxville, TN37996, USA }
\affiliation{          Joint Institute for Heavy Ion Research, \\
              P.O. Box 2008, MS6374, Oak Ridge, TN37831, USA}

\author{{\sc Ramon A. Wyss}}
\email{           wyss@kth.se}
\affiliation{          KTH (Royal Institute of Technology),\\
           AlbaNova University Centre,106 91 Stockholm}

\begin{abstract}

The response of isovector and isoscalar pairing to
generalized rotation in isospace is studied.
Analytical expressions for non-selfconsistent solutions
for different limiting cases of the model are derived. In particular,
the connections between gauge relations among pairing
gaps and the position of the isocranking axis are
investigated in $N$=$Z$ nuclei. The two domains
of collective and non-collective rotation in space
are generalized to isospace.  The amplitudes for
pair-transfer of $T=0$  and $T=1$ pairs are also
calculated. It is shown that the structure of
the $T=0$ state in odd-odd nuclei prevents any enhancement of
pairing transfer also in the presence of strong
$T=0$ correlations. The energy differences
of the $T=0$ and $T=1$ excitations in odd-odd nuclei are
qualitatively reproduced by Total Routhian Surface
calculations.
\end{abstract}


\maketitle

\newpage

\section{Introduction}\label{intro}

The event of detailed experimental studies of
heavy nuclei along the $N$=$Z$ line has resulted in
a revival of theoretical investigations related to
the properties of proton-neutron ($pn$) pairing
correlations~\cite{[Eng96],[Sat97],[Eng97],[Dea97],[Ter98],
[Dob98],[Fra99a],[Fra99],[Kan99],[Fra00],[Goo99],[Rop00]}.
In $N$=$Z$ nuclei one expects short range correlations
of $pn$ type to be of importance due to the four fold
degeneracy of the nuclear wave functions.
Of particular interest is the role played by
${\boldsymbol t}$=0 pairing correlations.  It is still
an issue of debate to what extent the correlations
in this channel of the pairing force are characterized
by a static gap similar to that of
the well established ${\boldsymbol t}$=1 seniority force.

The aim of our work is to devise a mean field model in which
we can describe consistently
excitations in real-  and iso-space on the same footing.
In a serie of papers we have shown that the isobaric analogue
states as well as ground state masses along the $N$=$Z$ line
form an unique probe to ${\boldsymbol t}$=0 pairing
correlations~\cite{[Sat97],[Sat00],[Sat01a],[Sat01b],[Sat01c]}.
Within the mean field approach,
the energy of isobaric analogue states can be described
by means of the isocranking approximation, analoguous to the
rotational excitations in real space~\cite{[Che78],[Sat01a]}.
The associated broken symmetry is the deformation of the pairing
field. The direction of the pairing field
vector is in turn intimately linked to the direction
of the cranking axis in isospace.
Hereafter (Sec.~\ref{t1} and~\ref{t0}) we derive analytical expressions
to elucidate these relations which, in fact, determine the regimes of
collective and non-collective rotations in
isospace.  These expressions and conclusions may be of
potential use in other, yet unknown, double-phase paired
systems that can be described by means of
an external cranking-type hamiltonian.

The collectivity of pairing
correlations can be acessed by means of pair transfer~\cite{[Broglia_book]}. 
In the analysis
of nuclei in the vicinity of the $N$=$Z$ line, it was concluded that indeed,
the ${\boldsymbol t}$=1, $pn$-pairing exhibits collectivity~\cite{[Bes77]}.
From this analysis, it was suggested that there is little evidence of
${\boldsymbol t}$=0 collectivity~\cite{[Bes00]}. Let us point out that
similar conclusions concerning the role of ${\boldsymbol t}$=0 pairing
has been drawn by Berkeley group~\cite{[Mac00s]} based on
the analysis of excitation energy spectra in $N$=$Z$ nuclei.
However, to draw definite conclusions a detailed understanding
of the structure of both $T=0$ and $T=1$ ground states in the odd-odd
(o-o) nuclei as well as the structure  even-even vacuum are necessary.
 There are empirical arguments based on isobaric symmetry
as well as theoretical arguments based essentially on time-reversal
symmetry breaking, which indicate that the
structure of the $T=1$ and $T=0$ states
in o-o $N$=$Z$ nuclei is entirely different~\cite{[Sat01b]}.
The simplest scenario, within the BCS theory, consistent with the data
can be reached by
assuming that the wave function of the  $T=0$ state in o-o nuclei
is a two quasi-particle excitation (2QP), whereas the
$T=0$ ground state in even-even (e-e) nuclei and the lowest $T=1$ state
in o-o nuclei are both local quasiparticle vaccua i.e. 0QP
states~\cite{[Sat01b]}. Within this interpretation
the transfer of a $T=0$ deuteron-like pair will always
be strongly quenched, irrespectively of the strength of the $T=0$
correlations, see Sec.~\ref{tran}. Hence, pair transfer may not necesserily
be a good indicator for the strength of $T=0$ pairing correlations.

The outline of the paper is the following:
In Sec.~\ref{t1} we derive analytical
solutions for the isocranked model with
${\boldsymbol t}$=1 pairing only. In Sec.~\ref{t0} we extend
these solutions to include also ${\boldsymbol t}$=0 pairing.
In Sec.~\ref{ssb} we discuss
the role of the isospin symmetry breaking mechanism due to
number projection and its influence on the Wigner energy.
In Sec.~\ref{tran} we discuss pair transfer from
$T=0$ and $T=1$ states in $N$=$Z$ o-o nuclei.
In Sec.~\ref{trs} we investigate the influence
${\boldsymbol t}$=0 pairing on nuclear
deformation by performing  Total Routhian Surface
(TRS) type calcualtions for
$T=0$ and $T=1$ states in $N$=$Z$ o-o nuclei.
In Sec.~\ref{isov} we discuss briefly
the shortcomings of our model due to the lack of
the particle-hole isovector field. We summarize
and conclude the paper in Sec.~\ref{sum}.

\section{Gauge invariance properties of proton-neutron coupled HFB
equations}\label{ggi}

The  HFB (BCS) equations:
\begin{equation} \left(
\ba{cc}
  {\boldsymbol h} & {\boldsymbol \Delta} \\
  -{\boldsymbol \Delta}^* & -{\boldsymbol h}^*
\ea \right)
\left( \ba{c}  {\boldsymbol U} \\ {\boldsymbol V} \ea \right)
=
\left( \ba{c}  {\boldsymbol U} \\ {\boldsymbol V} \ea \right) E
\end{equation}
are invariant under the transformation:
\begin{equation}
{\boldsymbol V} \longrightarrow e^{i\phi}{\boldsymbol V}
\end{equation}
which we call later global gauge invariance
transformation (GGI) since it requires both
proton  ${\boldsymbol V_p}
\rightarrow e^{i\phi} {\boldsymbol V_p}$ and neutron
${\boldsymbol V_n} \rightarrow e^{i\phi} {\boldsymbol V_n}$ amplitudes
to be transformed simoultanously. Indeed, in this case
the density matrix and pairing tensor transform as:
\begin{equation}
{\boldsymbol \rho} \equiv {\boldsymbol V}^* {\boldsymbol V}^T
\longrightarrow {\boldsymbol \rho}
\quad \mbox{and} \quad
{\boldsymbol \kappa} \equiv {\boldsymbol V}^* {\boldsymbol U}^T
\longrightarrow e^{-i\phi} {\boldsymbol \kappa}.
\end{equation}
Hence, the single-particle potential ${\boldsymbol \Gamma}\propto
{\boldsymbol \rho}$
and pairing potential  ${\boldsymbol \Delta}\propto {\boldsymbol \kappa}$
become
\begin{equation}
{\boldsymbol \Gamma} \longrightarrow {\boldsymbol \Gamma}
\quad \mbox{and} \quad
{\boldsymbol \Delta} \longrightarrow e^{-i\phi} {\boldsymbol \Delta}.
\end{equation}
The GGI allows, for example, to choose one of the pairing gaps to be
real and we will take advantage of it by assuming
[apart of Sec.~\ref{t0}] that the neutron gap $\Delta_{nn}$=$\Delta$$>$0.
However, in some cases and for sake of simplicity,
the eigenvectors will be given only up to the gauge transformation.

\section{Two-dimensional iso-cranking solutions of the $t=1$ pairing
model}\label{t1}

Let us consider in this section a model hamiltonian [$\hbar\equiv 1$
for convenience]:
\begin{equation}\label{ht1}
\hat H^{\omega_\tau} = \hat h_{sp} - G_{t=1}\hat {\boldsymbol P}_1^\dagger
\hat {\boldsymbol P}_1 -
\vec{\boldsymbol\omega}_\tau \vec{\hat {\boldsymbol t}},
\end{equation}
containing 
an iso-symmetric particle-hole
(ph) mean-field $\hat h_{sp}$. The  isovector ${\boldsymbol t}$=1
pairing interaction is generated by:
 \begin{equation}
 {\hat {\boldsymbol P}}_{1\pm 1}^\dagger  = \sum_{i>0} {\hat a}^\dagger_{in(p)}
 {\hat a}^\dagger_{\bar{i}n(p)} \quad\mbox{and}\quad
 {\hat {\boldsymbol P}}_{10}^\dagger  = \frac{1}{\sqrt2}
 \sum_{i>0} ( {\hat a}^\dagger_{in}{\hat a}^\dagger_{\bar{i}p} +
 {\hat a}^\dagger_{ip}{\hat a}^\dagger_{\bar{i}n} ),
 \end{equation}
 and  we consider two-dimensional iso-rotations
$\vec{{\boldsymbol\omega}}_\tau$=$[\omega_\tau \cos\varphi,
\omega_\tau\sin\varphi,0]$. Planar rotation in $N$=$Z$ nuclei are the most
general since  $\langle \hat t_z\rangle = 0$ due to number conservation.
We are interested in analytical but non-selfconsistent BCS solutions
within the constant gap approximation. Moreover, we assume
that $\lambda_n = \lambda_p = \lambda$ and
$|\Delta_{nn}| = \Delta = |\Delta_{pp}| $ but do not make any further
restrictions concerning the phase relations between
the gaps.
Within the BCS approximation and under the above mentioned
constraints the problem reduces
to a diagonalisation of a $4\times 4$ matrix with $\Delta$ and $\Delta_o$
being real [$\tilde e_i \equiv e_i -\lambda$] :
\begin{equation}\label{bcs}
\left[ \begin{array}{cc|cc}
\tilde e_i - E_i  &  - \frac{1}{2} \omega_\tau e^{-i\varphi} & \Delta
       &  e^{i\psi} \Delta_o \\ & & & \\
 - \frac{1}{2} \omega_\tau e^{i\varphi} & \tilde e_i - E_i
                        & e^{i\psi} \Delta_o  &
                                     e^{i\alpha} \Delta \\ & & & \\
\hline
 & & & \\
 \Delta    &  e^{-i\psi} \Delta_o & -\tilde e_i - E_i
               &  \frac{1}{2} \omega_\tau e^{i\varphi} \\ & & & \\
 e^{-i\psi} \Delta_o &  e^{-i\alpha} \Delta
                         & \frac{1}{2} \omega_\tau e^{-i\varphi} &
                                      -\tilde e_i - E_i
\end{array} \right]
\left[
\begin{array}{c}
U_{\bar i,n} \\ \\ U_{\bar i,p} \\ \\
\hline \\ V_{ i,n}
\\ \\ V_{ i,p} \\
\end{array} \right] = 0
\end{equation}
The physical (positive) eigenvalues of (\ref{bcs}) are:
\begin{equation}\label{qpr}
E_{i\pm} =
\sqrt{\tilde e_i^2 + \frac{1}{4} \omega_\tau^2 + \Delta_o^2 + \Delta^2 \pm
\sqrt X_i} \end{equation} where \begin{eqnarray} X_i = \tilde e_i^2
\omega_\tau^2 +
 4 \Delta^2\Delta_o^2 \cos^2\left(\psi -\frac{\alpha}{2} \right)
 +  \omega_\tau^2 \Delta^2 \sin^2\left(\varphi - \frac{\alpha}{2} \right)
& \nonumber \\
 - 4\tilde e_i\omega_\tau\Delta\Delta_o \cos
\left(\varphi - \frac{\alpha}{2} \right)
    \cos\left(\psi -\frac{\alpha}{2} \right) &
\end{eqnarray}
These roots are double-degenerated [Kramers degeneracy] with eigenvectors
of the form:
\begin{equation}\label{ev}
   \left[ \ba{c} U_{i\pm ,n} \\ U_{i\pm ,p}  \\
                 U_{\bar i\pm ,n} \\ U_{\bar i\pm ,p} \\
                 V_{i\pm ,n} \\ V_{i\pm ,p}  \\
                 V_{\bar i\pm ,n} \\ V_{\bar i\pm ,p}
          \ea \right]: \quad
  \longrightarrow \quad
   \left[ \ba{c} 0 \\ 0  \\
                 U_{\bar i\pm ,n} \\ U_{\bar i\pm ,p} \\
                 V_{i\pm ,n} \\ V_{i\pm ,p}  \\
                 0 \\ 0
          \ea \right]; \quad
   \left[ \ba{c} U_{\bar i\pm ,n} \\ U_{\bar i\pm ,p}  \\
                 0 \\ 0 \\
                 0 \\ 0  \\
                -V_{i\pm ,n} \\ -V_{i\pm ,p}
          \ea \right].
\end{equation}

\subsection{Model including $\boldsymbol t$=1, $\boldsymbol t_z$=$\pm 1$
pairing}\label{t1nn}


It is pedagogical to consider solutions to the model step by step
starting with the standard case of a $pn$ unpaired system $\Delta_o=0$.
In this case the eigenvalues reduce to:
\begin{equation}\label{rou}
E_{i\pm} = \sqrt{ \Delta^2 \cos^2\left(\varphi - \frac{\alpha}{2} \right) +
\left[ \frac{\omega_\tau}{2} \pm \sqrt{\tilde e_i^2 +  \Delta^2
\sin^2\left(\varphi - \frac{\alpha}{2}\right)}\right]^2}.
\end{equation}
Searching for the eigenvectors we assume that: ({\it i\/})
Satisfactory solutions must
obey {\it a minimal consistency condition\/}
and reproduce the initial [see eq. (\ref{bcs})]
relative phase relation for the gaps: $\Delta_{nn} = e^{i\alpha}\Delta_{pp}$.
({\it ii\/}) Moreover, we assume that neutron and
proton amplitudes can differ only by their phase i.e.
\begin{equation}\label{eq12}
U_{i,n}= e^{i\zeta} U_{i,p} \quad\mbox{and}\quad V_{i,n}=e^{i\eta}V_{i,p}
\end{equation}
which appears to be  a reasonable assumption especially for
systems described by common state independent order parameter $\Delta$. Under
these conditions we encounter only two types of solutions:
\begin{eqnarray}
& \displaystyle \varphi - \frac{\alpha}{2} =
                  ({2k+1})\frac{\pi}{2} \label{sol1}
\\
& \displaystyle \varphi - \frac{\alpha}{2} =
                    (2k)\frac{\pi}{2}     \label{sol2}
\end{eqnarray}

In the first case
$\varphi - \alpha/{2}=({2k+1}){\pi}/{2}$
the quasi-particle (qp) routhians (\ref{rou})
become linear as a function of iso-frequency:
\begin{equation}\label{rounc}
E_{j\pm} =
\left| E_j \pm \frac{1}{2} {\omega_\tau}\right|\quad \mbox{where} \quad
E_j = \sqrt{\tilde{e}_j^2 +  \Delta^2}.
\end{equation}
Before the first band crossing i.e. for frequencies lower
than
\begin{equation}
\omega_\tau \leq 2E_1 \equiv \omega_\tau^{(1)}
\end{equation}
where $E_1$ denotes energy of the lowest qp
at frequency zero, the qp routhians are
\begin{equation}
 E_{j\pm} = E_j \pm \frac{1}{2} {\omega_\tau}.
\end{equation}
The associated eigenvectors are also independent on
$\omega_\tau$ and equal:
\begin{equation}\label{evnc}
   \left[ \ba{c}
                 U_{\bar j\pm ,n} \\ U_{\bar j\pm ,p} \\
                 V_{j\pm ,n} \\ V_{j\pm ,p}  \\
          \ea \right]: \quad
  = \quad
   \left[ \ba{c} U_j \\ \mp e^{ i\varphi} U_j \\
                 V_j \\ \pm e^{-i\varphi} V_j
          \ea \right]
\end{equation}
where the $V_j$ and $U_j$ amplitudes are:
\begin{equation}\label{uvnc }
   V_j= \frac{1}{2} \sqrt{ 1- \frac{\tilde{e}_j}{E_j} }
                \quad\mbox{and}\quad
   U_j= \frac{1}{2} \frac{\Delta}{\sqrt{E_j( E_j -\tilde{e}_j)}}.
\end{equation}
For frequencies
$\omega_\tau^{(1)}\leq \omega_\tau \leq \omega_\tau^{(2)} \equiv 2E_2$
etc. standard procedure for blocked states can be applied to calculate
eigenvectors, see e.g. \cite{[Rin80]}.

The situation described here is characteristic for {\it non-collective\/}
type of rotation. Indeed, in this case the iso-alignment will change
stepwise at
each crossing frequency $\omega_\tau^{(i)}$ similar to the sp-model
described in detail in Refs.~\cite{[Sat01a],[Sat01b],[Sat01c]}.

\bigskip

The second possibility (\ref{sol2})
$\varphi - {\alpha}/{2}=k{\pi}$
yields routhians of standard BCS-type:
\begin{equation}\label{rouc}
E_{i\pm} = \sqrt{
\tilde e_{j\pm}^2  + \Delta^2 }.
\end{equation}
where $\tilde{e}_{j\pm} \equiv \tilde e_j \pm \frac{1}{2}{\omega_\tau}$.
The eigenvectors are:
\begin{equation}\label{evc}
   \left[ \ba{c}
                 U_{\bar j\pm ,n} \\ U_{\bar j\pm ,p} \\
                 V_{j\pm ,n} \\ V_{j\pm ,p}  \\
          \ea \right]: \quad
  = \quad
   \left[ \ba{r} U_{j\pm} \\ \mp e^{ i\varphi} U_{j\pm} \\
                 V_{j\pm} \\ \mp e^{-i\varphi} V_{j\pm}
          \ea \right]
\end{equation}
where
\begin{equation}
     V_{j\pm} = \frac{1}{2}\sqrt{ 1- \frac{\tilde{e}_{j\pm}}{E_{j\pm}} }
     \quad \mbox{and} \quad
      U_{j\pm} = \frac{1}{2}\frac{\Delta}
     {\sqrt{ E_{j\pm} (E_{j\pm} - \tilde{e}_{j\pm})}}
     \end{equation}
In this case the qp-routhians
have a nontrivial dependence on $\omega_\tau$.  They give rise to a smooth
alignment processes typical for {\it collective\/} rotation. The particular
case of this class of solutions corresponding to $\varphi=\alpha=0$ was
discussed in detail in ref.~\cite{[Sat01c]}.

\begin{figure}
\begin{center}
\leavevmode
\epsfysize=7.0cm
\epsfbox{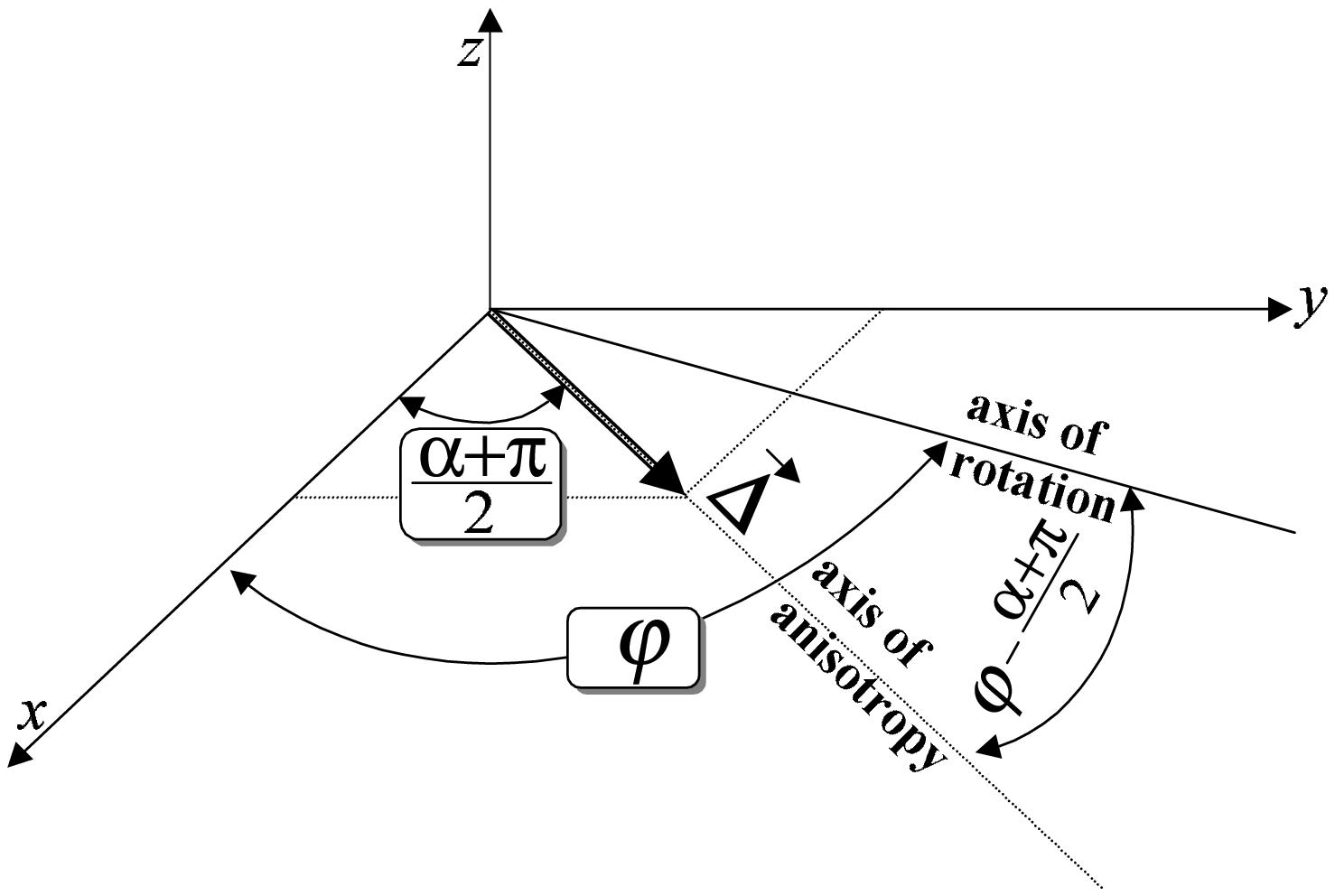}
\end{center}
\caption[]{
Schematic drawing showing the relative
position of the axis  of anisotropy [$\vec{\boldsymbol \Delta}$] versus
axis of isorotation.
The phase relations~\protect{(\ref{sol1})-(\ref{sol2})}
allow either for parallel [non-collective case] or perpendicular
[collective case] position of the rotation axes only. Note that
tilted solutions are
not allowed and hence,
three-dimensional isocranking in $N$=$Z$ nuclei can be effectively
reduced to one-dimensional theory.
}
\label{fig01}
\end{figure}

These two classes of solutions have a simple geometrical
interpretation+cite{[Gin68,Fra99]}.  Let us define the vector of anisotropy  in
isospace $\vec{\boldsymbol
\Delta} = [\Delta_x, \Delta_y, \Delta_z]$ as:
\begin{equation}
\Delta_x = \frac{1}{\sqrt2} (\Delta_{pp} - \Delta_{nn});\quad
\Delta_y = \frac{-i}{\sqrt2} (\Delta_{pp} + \Delta_{nn});\quad
\Delta_z = \Delta_{pn}
\end{equation}
In our case $\Delta_o =0$ and:
\begin{equation}
\frac{\Delta_x}{\Delta_y} = - \tan \left(\frac{\alpha}{2}\right).
\end{equation}
as shown schematically in fig.~\ref{fig01}.
Relations (\ref{sol1})-(\ref{sol2})
position the axis of iso-rotation  either parallel
or perpendicular with repect to $\vec{\boldsymbol \Delta}$
giving rise to non-collective or collective
iso-rotation, respectively. In particular, for most standard choices of phases
$\Delta_{nn} =\pm \Delta_{pp}$ the collective
axis is the $x(y)$-axis, respectively.
One may summarize that the possible solutions in this case only allow for
principal axis cranking, where the cranking axis is determined by the gauge
angle of the neutron and proton pair gap, respectively.

\subsection{Model including complete $\boldsymbol t$=1 pairing}\label{t1pn}

According to the geometrical interpretation of the pairing gaps,
switching on adiabatically $pn$ pairing $\Delta_{pn} [\equiv\Delta_z]$ should
always induce  {\it collectivity\/}. Indeed, in this case
 the qp routhians (\ref{qpr}) would depend on $\omega_\tau$ in
 a complicated, nonlinear
way for both cases (\ref{sol1}) and (\ref{sol2}).
Closer examination shows, however, that
the {\it non-collective\/} solution (\ref{evnc})
cannot be generalized to accomodate $pn$
pairing.~\footnote{At least not for a general case, see the discussion
on a similar subject in Sec.~\ref{t0t1}.}
On the contrary, {\it collective\/} solutions
(\ref{evc}) can be rather
strightforwardly extended to include $pn$ pairing.  In this case
 the qp routhians take the form:
 \begin{equation}\label{pnc}
E_{j\pm} = \sqrt{ \tilde{e}_{j\pm}^2 +
|\Delta\mp\Delta_{o}e^{i(\psi-\varphi)}|^2}.
\end{equation}
and the eigenvectors are:
\begin{equation}\label{eigc}
   \left[ \ba{c}
                 U_{\bar j\pm ,n} \\ U_{\bar j\pm ,p} \\
                 V_{j\pm ,n} \\ V_{j\pm ,p}  \\
          \ea \right]: \quad
  = \quad
   \left[ \ba{r} U_{j\pm} \\ \mp e^{ i\varphi} U_{j\pm} \\
                 V_{j\pm} \\ \mp e^{-i\varphi} V_{j\pm}
          \ea \right]
\end{equation}
where
\begin{equation}\label{uvt1}
     V_{j\pm} = \frac{1}{2}\sqrt{1-\frac{\tilde{e}_{j\pm}}{E_{j\pm}} }
     \quad \mbox{and} \quad
      U_{j\pm} = \frac{1}{2}\frac{\Delta\mp\Delta_{pn}e^{-i\varphi}}
     { \sqrt{ E_{j\pm} ( E_{j\pm} - \tilde{e}_{j\pm} )} }.
     \end{equation}

Let us observe that for $\psi-\varphi = (2n+1)\pi/2$ the routhians
(\ref{pnc}) have a similar form like  discussed
by Goodman~\cite{[Goo72]} for the static case of $\omega_\tau$=0:
\begin{equation}\label{goo}
E_{i\pm} =
\sqrt{ \tilde{e}_{i\pm}^2  + \Delta^2 + \Delta_o^2 }.
\end{equation}
For $\psi-\varphi =n\pi$,  on the other hand, one gets  routhians
similar in structure to those discussed by Bes {\it et al.}~\cite{[Bes00]}:
\begin{equation}\label{bes}
 E_{i\pm} = \sqrt{ \tilde{e}_{i\pm}^2
+ \left( \Delta \mp (-1)^n \Delta_o \right)^2 }.
\end{equation}

To gain further inside into our solutions let us calculate the
density matrix:
\beq
\rho_{i\tau,i\tau} =
\rho_{\bar i \tau,\bar i \tau} & = &
\frac{1}{4}\left\{2 -\frac{\tilde{e}_{i+}}{E_{i+}} -
\frac{\tilde{e}_{i-}}{E_{i-}} \right\} \quad\mbox{for} \quad \tau =n,p
\label{rhonn} \\
\rho_{in,ip} =
\rho_{\bar i n,\bar i p}  & = &
\frac{e^{-i\varphi}}{4}\left\{\frac{\tilde{e}_{i+}}{E_{i+}} -
\frac{\tilde{e}_{i-}}{E_{i-}} \right\} \label{rhonp}
\eeq
and pairing tensor:
\beq
\kappa_{jn,\bar jn} = e^{-i\alpha} \kappa_{jp,\bar jp} =
\frac{1}{4} \Delta X^{(+)}_j -
  \frac{1}{4} \Delta_o e^{i(\psi-\varphi)}X^{(-)}_j \\
\kappa_{in,\bar ip} = \kappa_{ip,\bar in} =
\frac{1}{4}\Delta_o e^{i\psi} X^{(+)}_j -
              \frac{1}{4}\Delta e^{i\varphi} X^{(-)}_j
\eeq
where
\begin{equation}\label{xpm}
X_i^{(\pm)} = \frac{1}{E_{i+}} \pm \frac{1}{E_{i-}}
\quad\mbox{and}\quad  X^{(\pm)} = \sum_i X_i^{(\pm)}
\end{equation}
Then, the gap equations are:
\beq
\frac{4}{G_{t=1 ,t_z=\pm 1}}  = X^{(+)} - \frac{\Delta_o}{\Delta}
e^{i(\psi -\varphi )} X^{(-)} \\
\frac{4}{G_{t=1,t_z=0}}  = X^{(+)} - \frac{\Delta}{\Delta_o}
e^{-i(\psi -\varphi )} X^{(-)}.
\eeq
For isospin symmetric pairing $G_{t=1 ,t_z=\pm 1}=G_{t=1 ,t_z=0}
=G_{t=1}$ the equations
can be solved either when: {\bf (a)} $X^{(-)}=0$ and for essentially
arbitrary values of gaps or {\bf (b)} for ${\Delta}={\Delta_o}$ and $\psi -
\varphi =n\pi$. The latter case leads to  gapless superconductivity
(\ref{bes}) with a rather unphysical qp-spectrum.

In particular, for the equidistant  $e_i = i\delta e$
level model with symmetric cut-off, the particle-hole symmetry
leads to  $X^{(-)}=0$ but only for the routhians of Goodman-type
(\ref{goo}). In such a case the particle
number condition $N=\mbox{Tr}\rho$ is also automatically satisfied
since [see eq. (\ref{rhonn})]:
\begin{equation}\label{phsym}
\sum_i \left\{\frac{\tilde{e}_{i+}}{E_{i+}} +
\frac{\tilde{e}_{i-}}{E_{i-}} \right\} \equiv 0.
\end{equation}
The alignment in iso-space is then given by the formula:
%
\be
T=\sqrt{\langle \hat t_x \rangle^2 + \langle \hat t_y \rangle^2 }
= \frac{1}{2}
\sum_{i>0} \left\{\frac{\tilde{e}_{i+}}{E_{i+}} -
\frac{\tilde{e}_{i-}}{E_{i-}} \right\}
\end{equation}
which is analogous to
the one obtained in ref.~\cite{[Sat01c]} [there, $\Delta^2$ should be
formally replaced by $\Delta^2 +\Delta_o^2$]. Moreover, since for
the isospin symmetric model $\Delta^2 + \Delta_o^2$ is constant the
moment of inertia in isospace (MoI-i) and all conclusions
drawn there remain unaffected.

For more complex routhians (e.g. given by eq.
(\ref{bes}) we have not been able find analytical
solutions even for such high-symmetry like the
equidistant level model.

\section{Models including isoscalar pairing.}\label{t0}

\subsection{Pure $\boldsymbol t$=0 pairing model.}\label{t0t0}

Let us now consider the model hamiltonian:
\begin{equation}\label{ht0}
\hat H^{\omega_\tau} = \hat h_{sp} - G_{t=1}\hat {\boldsymbol P}_1^\dagger
\hat {\boldsymbol P}_1
- G_{t=0}\hat {\boldsymbol P}_0^\dagger
\hat {\boldsymbol P}_0 -
\vec{\boldsymbol\omega}_\tau \vec{\hat {\boldsymbol t}},
\end{equation}
consisting of both isovector and isoscalar pairing interactions
coupling particles in time reversed orbits:
\begin{equation}
{\boldsymbol P}_{0}^\dagger  = \frac{1}{\sqrt2}
 \sum_{i>0} ( a^\dagger_{in}a^\dagger_{\bar{i}p} -
 a^\dagger_{ip}a^\dagger_{\bar{i}n} ).
 \end{equation}
Let us start our considerations with a pure
isoscalar $t=0$ model.
In this case the BCS equations take the following form [the pairing gap
$\Delta_0$ can be chosen real]:
\begin{equation}\label{bcst0}
\left[
\begin{array}{cc|cc}
\tilde{e}_i - E_i  &  - \frac{1}{2} \omega_\tau e^{-i\varphi} &
       &  -\Delta_0 \\ & & & \\
 - \frac{1}{2} \omega_\tau e^{i\varphi} & \tilde{e}_i - E_i
                        &  \Delta_0  &
                                       \\ & & & \\
\hline
 & & & \\
   &  \Delta_0^\ast & -\tilde{e}_i - E_i
               &  \frac{1}{2} \omega_\tau e^{i\varphi} \\ & & & \\
  -\Delta_0^\ast &
                         & \frac{1}{2} \omega_\tau e^{-i\varphi} &
                                      -\tilde{e}_i - E_i
\end{array} \right]
\left[
\begin{array}{c}
U_{\bar i,n} \\ \\ U_{\bar i,p} \\ \\
\hline \\ V_{ i,n}
\\ \\ V_{ i,p} \\
\end{array} \right] = 0
\end{equation}
The eigenvalues are linear in $\omega_\tau$ and equal:
\begin{equation}
E_{i\pm} = \left| \frac{1}{2}
{\omega_\tau} \pm E_i\right|
= E_i \pm \frac{1}{2} {\omega_\tau}
\quad \mbox{where} \quad  E_i = \sqrt{\tilde{e}_i^2 +  \Delta_0^2. }
\end{equation}
The equations are valid  below the first crossing frequency
 $\omega_\tau \leq \omega_\tau^{(1)} \equiv
 2 E_1$.
The roots are double-degenerated due to the Kramers degeneracy and
equal:
\begin{equation}\label{evt0}
   \left[ \ba{c} U_{i\pm ,n} \\ U_{i\pm ,p}  \\
                 U_{\bar i\pm ,n} \\ U_{\bar i\pm ,p} \\
                 V_{i\pm ,n} \\ V_{i\pm ,p}  \\
                 V_{\bar i\pm ,n} \\ V_{\bar i\pm ,p}
          \ea \right]:
  \longrightarrow
   \left[ \ba{c} 0 \\ 0  \\
                 U_{i} \\ \mp e^{i\varphi} U_{i} \\
                   V_{i} \\ \pm e^{-i\varphi} V_{i}  \\
                 0 \\ 0
          \ea \right],
   \left[ \ba{c} U_{i} \\ \mp e^{i\varphi} U_{i}  \\
                 0 \\ 0 \\
                 0 \\ 0  \\
                   V_{i} \\ \pm e^{-i\varphi} V_{i}
          \ea \right]
\end{equation}
where
\beq
 U_{i}=\frac{1}{2}
\sqrt{1 + \frac{\tilde{e}_i}{E_i}} \quad \mbox{and} \quad
 V_{i}=\frac{1}{2}
\sqrt{1 - \frac{\tilde{e}_i}{E_i}}
\eeq
independently on $\omega_\tau$.
The situation is similar to that
described by (\ref{rounc})-(\ref{evnc}) for the case of
$\boldsymbol t$=1, $\boldsymbol t_z$=$\pm 1$ pairing
in Sec.~\ref{t1nn}.
However, since the pair field is isotropic in isospace
only non-collective iso-rotation takes place regardless
of the direction of the iso-cranking axis.


\subsection{The $\boldsymbol t$=0 plus  $\boldsymbol t$=1,
$\boldsymbol t_z$=$\pm 1$ pairing model.}\label{t0t1}


The extension of the pure $\boldsymbol t$=0 model including
$\boldsymbol t$=1, $\boldsymbol t_z$=$\pm 1$ can be done by either
linking it smoothly to {\it non-collective\/} solutions
(\ref{evnc}) or {\it collective\/} solutions (\ref{evc}).
Let us consider first {\it non-collective\/}
solutions obeying the condition (\ref{sol1}). These can be
relatively easy generalized to include $\boldsymbol t$=0 pairing
provided that the following phase relation is satisfied:
\begin{equation}
\theta -\frac{1}{2}\alpha = n\pi
\end{equation}
where $\Delta_0 = e^{i\theta}|\Delta_0|$. This is a necessary condition
for the linear term of the determinant of the BCS matrix to vanish and, in turn,
to preserve the mirror symmetric structure [$E_i,-E_i$] of the HFB solutions.
In this case our solutions have {\it non-collective\/} character.
The quasiparticle
routhians below the first crossing frequency take the following form:
\begin{equation}
E_{i\pm} = E_i \pm \frac{1}{2}\omega_\tau\quad \mbox{where}
\quad  E_i = \sqrt{\tilde{e}_i^2 + \Delta^2 + |\Delta_0|^2}
\end{equation}
The associated eigenvectors are:
\begin{equation}\label{evt0t1a}
   \left[ \ba{c} U_{i\pm ,n} \\ U_{i\pm ,p}  \\
                 U_{\bar i\pm ,n} \\ U_{\bar i\pm ,p} \\
                 V_{i\pm ,n} \\ V_{i\pm ,p}  \\
                 V_{\bar i\pm ,n} \\ V_{\bar i\pm ,p}
          \ea \right]:
  \longrightarrow
   \left[ \ba{c} 0 \\ 0  \\
                 U_{i\pm} \\ \mp e^{i\varphi} U_{i\pm} \\
                 V_{i\pm} \\  \pm e^{-i\varphi} V_{i\pm}  \\
                 0 \\ 0
          \ea \right],
   \left[ \ba{c} - U^*_{i\pm} \\ \pm e^{i\varphi} U^*_{i\pm}  \\
                 0 \\ 0 \\
                 0 \\ 0  \\
                 V_{i\pm} \\ \pm e^{-i\varphi} V_{i\pm}
          \ea \right].
 \end{equation}
 where
\begin{equation}
V_{i\pm} = \frac{1}{2} \sqrt{1-\frac{\tilde{e}_i}{E_i}}; \quad \mbox{and}
\quad
U_{i\pm} = \frac{1}{2} \frac{ \Delta\mp  e^{-i\varphi}\Delta_0}
                             {\sqrt{E_i (E_i -\tilde{e}_i)}}.
\end{equation}
However, since the pairing tensors are:
\begin{equation}
\kappa_{in,\bar{i}n} = e^{-i\alpha} \kappa_{ip,\bar{i}p} =
\frac{\Delta}{2E_i}\quad\mbox{and}\quad
\kappa_{in,\bar{i}p} = - \kappa_{ip,\bar{i}n}  =
\frac{\Delta_0}{2E_i}
\end{equation}
the selfconsitency conditions for pairing gaps yield the additional constraint:
\begin{equation}
\frac{2}{G_{t=1}} = \sum_{i>0} \frac{1}{E_i} =  \frac{2}{G_{t=0}}
\end{equation}
for the coupling constants.  Solutions are therefore possible
only for $G_{t=1}$=$G_{t=0}$.
Coexistance of this type was already reported in the
literature~\cite{[Sat97],[Eng96],[Eng97]}.

\bigskip

Generalization of the {\it collective\/}
solution (\ref{sol1}) to include $\boldsymbol t$=0 pairing
is far more difficult. It can be shown,
however, that for the following phase relations:
\begin{equation}
\theta -\varphi = (2n+1)\frac{\pi}{2},
\end{equation}
the eigenvectors take the following form:
\begin{equation}\label{evt0t1b}
   \left[ \ba{c} U_{i\pm ,n} \\ U_{i\pm ,p}  \\
                 U_{\bar i\pm ,n} \\ U_{\bar i\pm ,p} \\
                 V_{i\pm ,n} \\ V_{i\pm ,p}  \\
                 V_{\bar i\pm ,n} \\ V_{\bar i\pm ,p}
          \ea \right]:
  \longrightarrow
   \left[ \ba{c} 0 \\ 0  \\
                 U_{i\pm} \\ \mp e^{i\varphi} U_{i\pm}^* \\
                 V_{i\pm} \\  \mp e^{-i\varphi} V_{i\pm}^*  \\
                 0 \\ 0
          \ea \right],
   \left[ \ba{c} - U^*_{i\pm} \\ \pm e^{i\varphi} U_{i\pm}  \\
                 0 \\ 0 \\
                 0 \\ 0  \\
                 V_{i\pm}^* \\ \pm e^{-i\varphi} V_{i\pm}
          \ea \right].
 \end{equation}
 In this case the quasiparticle routhians are:
\begin{equation}
E_{i\pm} =
\left( \tilde{e}_{i}^2 + \frac{\omega_\tau^2}{4}
+\Delta^2 +  |\Delta_0|^2 \pm 2 \sqrt{
 \frac{\omega_\tau^2}{4} (\tilde{e}_{i}^2 + |\Delta_0|^2 )
 + \Delta^2 |\Delta_0|^2 } \right)^{1/2}
\end{equation}

For the general case of $\omega_\tau \ne 0$ the amplitudes are
 given by lenghty and rather non-transparent expressions.
 However, the property  of these solutions are  easily recognized
already at $\omega_\tau = 0$ when:
\begin{eqnarray}
V_{i\pm} = \frac{e^{-i\frac{(\varphi-\theta)}{2}}}{\sqrt2}
\sqrt{1-\frac{\tilde{e}_i}{E_{i\pm}}}
\cos\frac{\varphi-\theta}{2}  \\
U_{i\pm} = \frac{e^{-i\frac{(\varphi-\theta)}{2}}}{\sqrt2}
 \frac{ \Delta\pm |\Delta_0| }
{\sqrt{E_{i\pm} (E_{i\pm} -\tilde{e}_i)}}
\cos\frac{\varphi-\theta}{2}
\end{eqnarray}
and
\begin{equation}
E_{i\pm} = \sqrt{\tilde{e}_i^2 + (\Delta \pm |\Delta_0|)^2 }.
\end{equation}
The gap equations take the following form:
\begin{eqnarray}
\frac{4}{G_{t=1}}  = X^{(+)} + \frac{|\Delta_0|}{\Delta} X^{(-)} \\
\frac{4}{G_{t=0}}  = X^{(+)} + \frac{\Delta}{|\Delta_0|} X^{(-)}
\end{eqnarray}
where $X^{(\pm )}$ are defined as in eq.~(\ref{xpm}). Since
$X^{(-)}<0$ the gap equations can be solved only
when $G_{t=0}$=$G_{t=1}$ and $\Delta =|\Delta_0|$. In this case
the quasiparticle spectrum takes the rather unphysical gapless form
in spite of the strength of pairing force. Let us mention here
that particle-number is in this case automatically satisfied
at least for equidistant spectrum since in this case
particle-hole symmetry [eq.~(\ref{phsym})] is fulfilled.

\smallskip

\section{Isospin symmetry breaking}\label{ssb}

\smallskip

The considerations of Sec.~\ref{t1} show that
{\it collective\/} iso-rotations  are possible when
the isovector pairing field $\vec{{\boldsymbol \Delta}}$ is
perpendicular to the axis of iso-cranking.  Such a collective
motion allows to take into account isospin fluctuations
in a static way by replacing the cranking constraint:
\begin{equation}\label{ct}
\sqrt{\langle \hat t_x \rangle^2 + \langle \hat t_y \rangle^2 } = T
\end{equation}
by
\begin{equation}\label{ctt}
 \sqrt{\langle \hat t_x
\rangle^2 + \langle \hat t_y \rangle^2 } =\sqrt{T(T+1)}
\end{equation}
In this way the Wigner energy [$E_W\sim T$] can be, at least
partially, restored.
Quantitative calculations
\cite{[Sat01a],[Sat01b],[Sat01c]}
show, however, that in the presence of the standard $\boldsymbol t$=1
field, the MoI is too large and cannot account for the empirical data.
In these calculations the isovector part of the {\it particle-hole\/}
field is not taken into account~\cite{[Nee02]} (see also
Sec.~\ref{isov}). In the absence of these correlations
the mechanism which lowers the MoI in isospace was fully
ascribed to isoscalar pairing in complete
analogy to the mechanism of lowering the spatial MoI by
the isovector superfluidity.

However, as demonstrated in Sect.~\ref{t0},
at the level of the BCS approximation one cannot
obtain a mixed {\it collective\/} solution which in turn is necessary to
lower the MoI and to account for isospin fluctuations in a
static way (\ref{ctt}). To make use of the static condition
(\ref{ctt}) and to restore (locally) the MoI in isospace it
is necessary to find phase mixed solutions.
It has already been shown~\cite{[Sat97]} that approximate number
projection of Lipkin-Nogami type allows
for the mixing of ${\boldsymbol t}$=1 and
${\boldsymbol t}$=0 pairing
phases, breaking the isospin symmetry.
The mixing appears only when the strength
$G_{t=0}$ of the average isoscalar field
exceeds the strength of the isovector pair field
$G_{t=1}$ i.e. for $x^{t=0}=G_{t=0}/G_{t=1}\geq x_{crit}\sim 1.1$
\footnote{Numerical estimate for $N=Z$ nuclei are given in Ref.~\cite{[Sat00]}.}
The Wigner energy generated through this mechanism
contributes with opposite sign to
$\lambda_{\tau\, \tau}^{(2)}$ and $\lambda_{\tau\, -\tau}^{(2)}$
building up an asymmetry in the auxilliary LN fields:
\begin{equation}
\lambda_{\tau \tau\prime}^{(2)} \Delta N_\tau
\Delta N_{\tau\prime}.
\end{equation}

Since the LN solution does not allow for an isovector
$pn$-field, the situation qualitatively resembles
the case considered in Sec.~\ref{t0t1}. Let us
however point out that the LN procedure introduces
an explicit state dependence of the effective gaps
\begin{equation}
\Delta_{i\tau \bar{j}\tau\prime} =
\Delta_{\tau\tau\prime}\delta_{ij}+2\lambda_{\tau \tau}^{(2)}
\kappa_{i\tau\bar{j}\tau\prime}.
\end{equation}
Moreover, the Lipkin Nogami parameters $\lambda_{\tau \tau\prime}^{(2)}$
are anisotropic in isospace. In turn, a wider class of solutions
become possible, including solutions which mix
$\boldsymbol t$=1 $\boldsymbol t$=0 pairing phases
and are {\it collective\/} in isospace.
The absence of the isovector $pn$-field in the LN solutions limits,
however, the possibilities to calculate the amplitude of
the isovector $pn$-pair transfer, see next section.


\section{Proton-neutron pair transfer}\label{tran}


The theory of pair transfer/stripping processes
like ($\alpha$,d),($^3$He,n) etc. were developed already in the end
of 50's and the beginning of 60's
\cite{[ElN57],[ElN59],[New60],[Gle62]}
based on the plane wave Born approximation.
The general expression for two-nucleon spectroscopic factor (2nSF)
carrying nuclear structure information was derived using
the spherical shell model in the $jj-$coupling limit,
see Refs.~\cite{[New60],[Gle62]}.
The first analysis of the 2nSF using the pairing interaction
model was done by Yoshida~\cite{[Yos62]}. He pointed out
the possibility of an enhanced cross-section
for the two-particle transfer of the isovector pair due to
collective pairing phenomena which may be
particularly strong if the pairing coherency extends over
few $j-$shells. Based on Yoshida's work, Fr\"obrich
\cite{[Fro70],[Fro71]} has analysed the influence
of $pn$ pairing on $pn$-pair transfer in
$N$=$Z$ nuclei using
both $\boldsymbol t$=1 and $\boldsymbol t$=0 pairing interactions
within a single-$j$ shell model space. He pointed
out that $pn$-pairing can enhance the cross-section
by a factor of 3 as compared to conventional
shell-model calculations of~\cite{[Fle68]}.

\begin{figure}
\begin{center}
\leavevmode
\epsfysize=7.0cm
\epsfbox{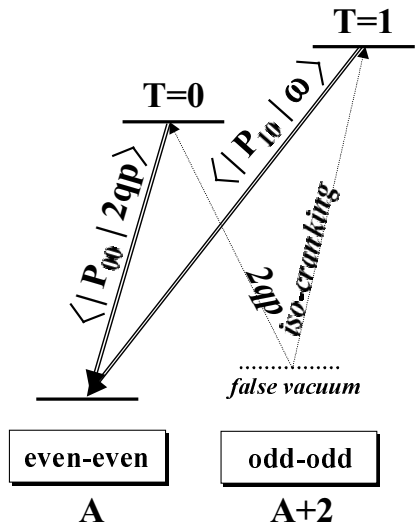}
\end{center}
\caption[]{
Schematic figure of the pair transfer and structure
of the T=0 and T=1 states in o-o ($A+2$) and ground state
of e-e ($A$) $N$=$Z$ nuclei. The thin arrows indicate
the structure of the T=0(1) states in o-o nuclei
in our model. Note, that the different structures of the
T=0 states imply
the quenching of isoscalar pair transfer even for
a $\boldsymbol t$=0 paired systems.
}
\label{fig02}
\end{figure}

In the deformed shell model [paired mean-field] the
differential cross-section
describing the transfer of a structureless isoscalar or
isovector $pn$ pair is proportional to:
\begin{equation}\label{amp}
\frac{\mbox{d}\sigma}{\mbox{d}\Omega} \sim
|T^{(A,A+2)}_{10(00)}|^2 = | \langle \Psi_A | {\boldsymbol P}_{10(00)} |
 \Psi_{A+2}(T) \rangle |^2
\end{equation}
where we have assumed that the transfer goes from
the $T=1(0)$ ground state  $|\Psi_{A+2}(T)\rangle$ of the o-o
$N$=$Z$=$(A+2)/2$ nucleus to the ground state $|\Psi_A\rangle$
of the even-even nucleus $N$=$Z$=$A/2$.
Within our model~\cite{[Sat01a],[Sat01b],[Sat01c]}
$|\Psi_A\rangle$ has the
structure of a 0qp state while the structure of
$|\Psi_{A+2}(T)\rangle$ depends on the isospin $T$. For $T$=0,
$|\Psi_{A+2}(T=0)\rangle$ is a two-quasiparticle state while
the $|\Psi_{A+2}(T=1)\rangle$ states maintain the $0qp$
structure but is cranked in isospace, see fig.~\ref{fig02}.
In all cases the $|\Psi_A\rangle$ and
$|\Psi_{A+2}(T)\rangle$ states are
described by the fully self-consistent
amplitudes (${\boldsymbol U}^{(A)}\, {\boldsymbol V}^{(A)}$) and
(${\boldsymbol U}^{(A+2)}\, {\boldsymbol V}^{(A+2)}$).
Therefore, one can make use of
the generalized Wick's theorem to
derive the explicit expression for the
pair transfer amplitude  $T^{(A,A+2)}$
(\ref{amp}):
\begin{equation}\label{ampeq}
 T^{(A,A+2)}_{10(00)}
  =  \frac{1}{\sqrt 2} \langle \Psi_A | \Psi_{A+2} \rangle
       \sum_{i>0} \left\{ \kappa^{(A,A+2)}_{in,\bar{i}p} \pm
        \kappa^{(A,A+2)}_{ip,\bar{i}n} \right\}
\end{equation}
where
\beq
  {\boldsymbol \kappa}^{(A,A+2)} =
         ({\boldsymbol V}^{(A+2)})^\ast  ({\boldsymbol U}^{T})^{-1}
                 ({\boldsymbol U}^{(A)})^T  \\  
      {\boldsymbol U} \equiv ({\boldsymbol U}^{(A+2)})^\dagger
                             {\boldsymbol U}^{(A)} +
      ({\boldsymbol V}^{(A+2)})^\dagger {\boldsymbol V}^{(A)}.
\eeq
The overlap is given by Onishi formula:
\begin{equation}
  \langle \Psi_A | \Psi_{A+2} \rangle  = \sqrt{\mbox{Det}{\boldsymbol U}}.
\end{equation}

 According to our model there is a fundamental difference between
 the structure of the $T$=0 and $T$=1 states in the parent [o-o]
 nucleus, see fig.~\ref{fig02}. The $T$=0 states correspond to
 2QP excitations.
 Therefore the deuteron will be transferred from a 2QP
 state to the BCS vacuum in the daughter nucleus. In such a case
 the  $T^{(A,A+2)}_{00}$ amplitude  will essentially probe
 $\sim (U_A)^2_i$  [$i$ enumerates blocked $qp$ state] and
 will always be severly quenchend~\cite{[Yos62]}.

\begin{figure}
\begin{center}
\leavevmode
\epsfysize=10.0cm
\epsfbox{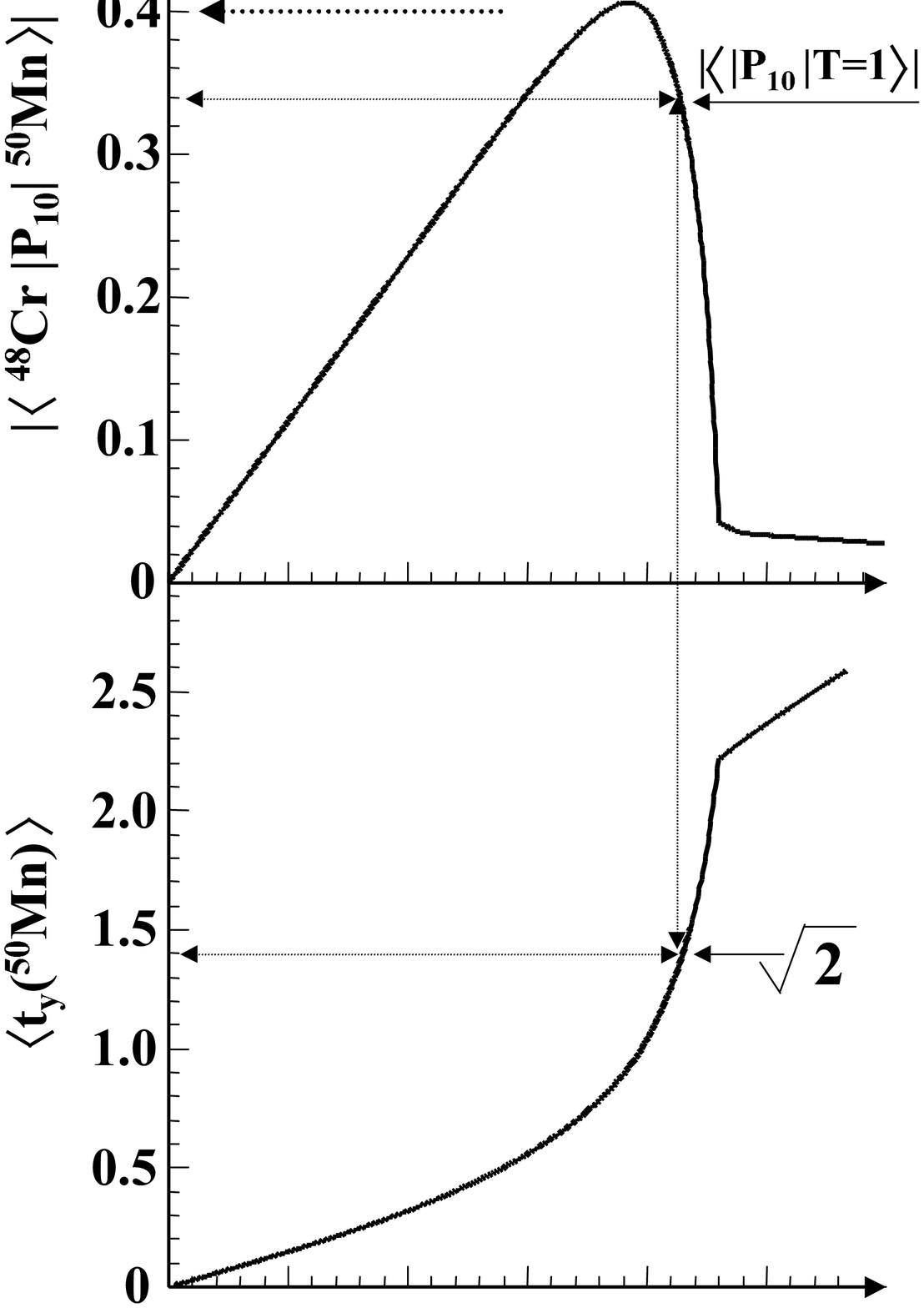}
\end{center}
\caption[]{
Isovector $pn$-pair transfer
$\langle | P_{10} | \omega_\tau \rangle$ [upper part]
and   isoalignment [lower part] versus $\omega_\tau$.
The calculations were done in  the LN approximation. In this
approximation $\boldsymbol t$=1 $pn$ pairing plays a
redundant role~\cite{[Sat97],[Sat00]}
resulting in the quenching of
$\langle | P_{10} | \omega_\tau \rangle$.
}
\label{fig03}
\end{figure}

 On the other hand the isovector $pn$-pair will be transfered between
 $0qp$ states and can, in principle, be enhanced due to
 the pairing collectivity.  The generic situation
 is illustrated in fig.~\ref{fig03} for
 the case of $^{48}$Cr and  $^{50}$Mn. In these calculations we assume
 $x^{t=0} =1.3$ for both the ground state of $^{48}$Cr and
 the {\it false vacuum\/} of $^{50}$Mn. With increasing $\omega_\tau$
 $^{50}$Mn picks up a small fraction of the isovector $pn$ pairing and
 $T^{(A,A+2)}_{10}$ increases reaching maximum around $T_x\sim 1$.
 Then in the region of the phase transition it decreases rapidly
 due to the vanishing overlap, see eq.~(\ref{ampeq}).
 In the region of $T_x\sim$1 to $\sim\sqrt2$
 the  $|T^{(A,A+2)}_{10}| \sim |T^{(A,A+2)}_{00}| \sim 0.4$
 i.e. no enhancement is calculated for the pair transfer.

\begin{figure}
\begin{center}
\leavevmode
\epsfysize=7.0cm
\epsfbox{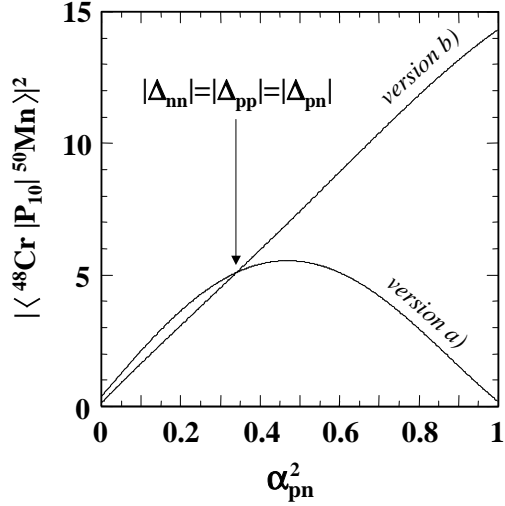}
\end{center}
\caption[]{
Pair transfer amplitude
$\langle | P_{10} | \rangle$
calculated in the BCS approximation for a model
involving $\boldsymbol t$=1 correlations
only.
The results are shown as a function
of the contribution of $\Delta_{pn}$
to the total gap, see text for details.
}
\label{fig04}
\end{figure}

 The quenching of the $T^{(A,A+2)}_{10}$ amplitude
 is related to the fact that within the $LN$ model the isovector
 $pn$ pairing plays essentially a redundant role.
 To investigate the importance of isovector $pn$ correlations
 we have to come back to the $\boldsymbol t$=1 pairing model of
 Sec.~\ref{t1pn}. In this case an analytical expression for
 the transfer amplitude can be easily derived:
\begin{equation}\label{ampbcs}
 \frac{T^{(A,A+2)}_{10}}{ \langle \Psi_A | \Psi_{A+2} \rangle  }
    = \frac{e^{i\varphi}}{\sqrt 2}
    \sum_{i>0} \left\{
          \frac{{u_{i+}^{(A)}}^\ast  v^{(A+2)}_{i+}}{X^{(-)}_i}
       -  \frac{{u_{i-}^{(A)}}^\ast  v^{(A+2)}_{i-}}{X^{(+)}_i}
             \right\}
\end{equation}
where
\begin{equation}
X^{(\pm )}_i = {u_{i\pm}^{(A)}}^\ast u^{(A+2)}_{i\pm}
  + {v_{i\pm}^{(A)}}^\ast v^{(A+2)}_{i\pm}.
\end{equation}
The amplitudes (${\boldsymbol u}, {\boldsymbol v}$) define
the Bogolyubov transformation in
the canonical basis:
\beq
|\pm\rangle = \frac{1}{\sqrt 2} ( |n\rangle \pm e^{i\varphi} |p\rangle ) \\
\widetilde{|\pm}\rangle = \frac{1}{\sqrt 2}
( \overline{|n}\rangle \pm e^{i\varphi} \overline{|p}\rangle ).
\eeq
They are equal:
\begin{equation}\label{uv}
     u_{i\pm} = \sqrt{2} U_{i\pm}^\ast \quad
\mbox{and} \quad
     v_{i\pm} = \sqrt{2} V_{i\pm}^\ast
\end{equation}
where $U_{i\pm},V_{i\pm}$ are given by eq.~(\ref{uvt1})
[for the daughter nucleus they are obtained at $\omega_\tau$=0].
Let us now assume that $\omega_\tau$=0. When $\Delta_{pn}$=0 then
both $u_+ = u_-$ and $v_+ = v_-$ and we have an exact cancellation
of two relatively large terms in the r.h.s of eq.~(\ref{ampbcs}).
On the other
hand, for $\Delta_{pn}\ne$0, also $u_+ \ne u_-$ and we
have a positive interference originating entirely from the terms
proportional to $\Delta_{pn}$ and, in turn, a rapid increase
of $|T^{(A,A+2)}|$ as a function of the contribution of the $pn$
pair gap
to the total gap
$\Delta_T^2 \equiv \Delta^2 + |\Delta_{pn}|^2=const$. This is shown
in fig.~\ref{fig04} in version a)
In this version of the calculations
$\alpha_{pn}^2$ [$|\Delta_{pn}|^2 \equiv \alpha_{pn}^2 \Delta_T^2$]
was forced to be the same for $^{48}$Cr and  $^{50}$Mn
(at $\omega_\tau =0$).  Version b),
on the other hand, assumes fixed structure $\Delta = |\Delta_{pn}|$
in $^{48}$Cr. In this case $\alpha_{pn}^2$ refers to $^{50}$Mn.
It is clearly seen from the figure that the maximum transfer
is expected for similar [but no necesserly equal] content
of $pn$ pairing in the parent and daughter nucleus. Obviously, other
factors like different deformations etc. may additionally hinder
$pn$-transfer.

\section{Total Routhian Surface calculations including isoscalar
pairing}\label{trs}

We have performed systematic calculations of the energy differences
$\Delta E_T \equiv E_{T=0} - E_{T=1}$ between the lowest
$T=0$ and $T=1$ states in o-o $N$=$Z$ nuclei using the
Total Routhian Surface (TRS) method. We used
the deformed Woods-Saxon potential
and included both isovector
and isoscalar pairing treated in
the LN approximation.
The deformation space covered quadrupole $\beta_2,\,\gamma$
and hexadecapole $\beta_4$ shapes. The
liquid-drop formula
of~\cite{[Mye66]} was used to calculate the macroscopic part
of the total energy. Apart of
including $\boldsymbol t$=0 pairing, the details of the
method follow our standard implementation and we refere the
reader to Ref.~\cite{[Sat94]} for further details.

\begin{table}
\begin{center}
\begin{tabular}{crrrrr}
\hline
\hline
     &   $E^{T=0}_1$  & $E^{T=0}_2$  &
  $E^{T=1}_1$  & $E^{T=1}_2$ \\
\hline
\hline
 $^{6}_{3}\mbox{Li}_{3}$    &       0 ($1^+$)  &    2.185 ($3^+$)
                            &   3.562 ($0^+$)  &    5.370 ($2^+$)  \\

 $^{10}_{5}\mbox{B}_{5}$    &       0 ($3^+$)  &    0.718 ($1^+$)
                            &   1.742 ($0^+$)  &    5.163 ($2^+$) \\

 $^{14}_{7}\mbox{N}_{7}$    &       0 ($1^+$)  &    4.915 ($0^-$)
                            &   2.313 ($0^+$)  &    8.062 ($1^-$) \\

 $^{18}_{9}\mbox{F}_{9}$    &       0 ($1^+$)  &    0.937 ($3^+$)
                            &   1.041 ($0^+$)  &    3.061 ($2^+$) \\

 $^{22}_{11}\mbox{Na}_{11}$ &       0 ($3^+$)  &    0.583 ($1^+$)
                            &   0.657 ($0^+$)  &    1.952 ($2^+$) \\

 $^{26}_{13}\mbox{Al}_{13}$ &       0 ($5^+$)  &    0.417 ($3^+$)
                            &   0.228 ($0^+$)  &    2.070 ($2^+$) \\

 $^{30}_{15}\mbox{P}_{15}$  &       0 ($1^+$)  &    0.709 ($1^+$)
                            &   0.677 ($0^+$)  &    2.938 ($2^+$) \\

 $^{34}_{17}\mbox{Cl}_{17}$ &   0.146 ($3^+$)  &    0.461 ($1^+$)
                            &       0 ($0^+$)  &    2.158 ($2^+$) \\

 $^{38}_{19}\mbox{K}_{19}$  &       0 ($3^+$)  &    0.459 ($1^+$)
                            &   0.130 ($0^+$)  &    2.403 ($2^+$) \\

 $^{42}_{21}\mbox{Sc}_{21}$ &   0.611 ($1^+$)  &    0.617 ($7^+$)
                            &       0 ($0^+$)  &    1.586 ($2^+$) \\

 $^{46}_{23}\mbox{V}_{23}$  &   0.801 ($3^+$)  &    0.915 ($1^+$)
                            &       0 ($0^+$)  &    0.993 ($2^+$) \\

 $^{50}_{25}\mbox{Mn}_{25}$ &   0.230 ($5^+$)  &    0.651 ($1^+$)
                            &       0 ($0^+$)  &    0.800 ($2^+$) \\

 $^{54}_{27}\mbox{Co}_{27}$ &   0.199 ($7^+$)  &    0.937 ($1^+$)
                            &       0 ($0^+$)  &    1.447 ($2^+$) \\

 $^{58}_{29}\mbox{Cu}_{29}$ &       0 ($1^+$)  &    0.445 ($1^+$)
                            &   0.203 ($0^+$)  &    1.652 ($2^+$) \\

 $^{62}_{31}\mbox{Ga}_{31}$ &   0.571 ($1^+$)  &    0.818 ($3^+$)
                            &       0 ($0^+$)  &                  \\

 $^{66}_{33}\mbox{As}_{33}$ &   0.837 [$1^+$]  &    1.231 ($3^+$)
                            &       0 ($0^+$)  &    0.963 [$2^+$] \\

 $^{70}_{35}\mbox{Br}_{35}$ &   1.337 ($3^+$)  &    1.653 ($5^+$)
                            &       0 ($0^+$)  &    0.934 ($2^+$)\\

 $^{74}_{37}\mbox{Rb}_{37}$ &   1.006 [$3^+$]  &  1.224 [$4^+$]
                            &       0 [$0^+$]  &    0.478 [$2^+$] \\
\hline
\hline
\end{tabular}
\caption[A]{
The lowest $T=0$ and $T=1$ states in odd-odd N=Z nuclei.
Data are taken from:  $^{18}F$ -- $^{58}Cu$~Ref.~\cite{[Fir96]};
$^{62}Ga$~Ref.~\cite{[Vin98]};
$^{66}As$~Ref.~\cite{[Grz01]};  $^{70}Br$~Ref.~\cite{[Jen02]};
$^{74}Kr$~Ref.~\cite{[Rud96]}. Spin values in square
parantheses are uncertain.}
\label{T1}
\end{center}
\end{table}

The calculations have been done for all o-o $N$=$Z$ nuclei from
$^{18}$F to $^{74}$Rb, see table~\ref{T1}.
To account for the $T=0$ states
the 2QP surfaces $\alpha^\dagger_1 \alpha^\dagger_2 |0\rangle$
have been created where $\alpha^\dagger_{1(2)}$ denote the lowest
qp excitations of the same signature. We use the notion of signature
rather than time-reversal since the calculations have been done at
a small fixed spatial rotational frequency. By blocking
$\alpha^\dagger_1 \alpha^\dagger_2 |0\rangle$
we assure that all pairing fields are truely blocked.
Indeed, blocking the lowest qp states of different signatures
$\alpha^\dagger_1 \alpha^\dagger_{\tilde 1} |0\rangle$ results
in the blocking of isovector $pp$ and $nn$ pairing but
not isoscalar $pn$ pairing.  It is assumed that the valence blocked
quasiparticles [proton-neutron] may interact. This interaction
is added as a perturbation assuming a simple [isoscalar] delta
force $g(A)_{eff}\delta({\boldsymbol r}_1 -
{\boldsymbol r}_2)$ as a residual interaction.
For further simplification we assume either
spherical $|(nljm)^2;I\rangle$ or Nilsson
asymptotic $| (Nn_z\Lambda K)^2 \rangle$ limits
for the two-body wave function with $I=I_{exp}$ or
$I_{exp}=2K$, respectively.
An effective strength of the
residual force $g(A)_{eff} = g_{eff}/\sqrt{A}$
was assumed.  This gives rise to a
$\sim 1/A$  dependence for the matrix element in accordance to
the standard mass dependence of the
valence $pn$ interaction in liquid-drop
models~\cite{[Kra79]}.

The $T=1$ states were calculated in a slightly simplified manner.
Self-consistent TRS calculations have been done only for the
0QP state [{\it false vacuum\/}] at $\omega_\tau$=0. Then, the
isocranking calculations were performed at
the deformation minimum of the TRS calculations.

\begin{figure}
\begin{center}
\leavevmode
\epsfysize=10.0cm
\epsfbox{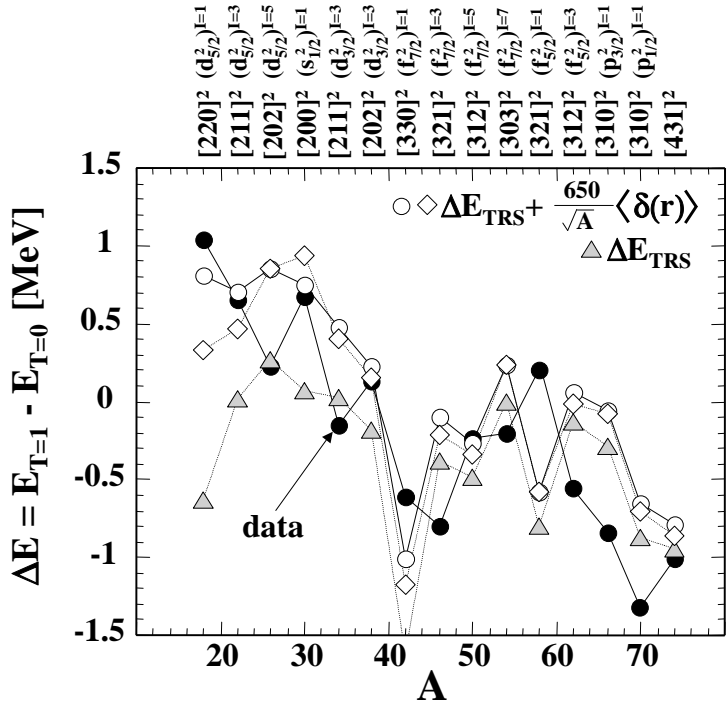}
\end{center}
\caption[]{
The difference  $\Delta E_T \equiv E_{T=0} - E_{T=1}$ calculated
using the TRS model [triangles]. Open circles and diamonds mark
the calculations including the correction due to
the interaction of the valence particles
evaluated assuming Nilsson asymptotic and spherical wave
functions, respectively [see text]. Solid circles mark
the available data which are collected in Table~\protect{\ref{T1}}.
}
\label{fig05}
\end{figure}

The TRS calculations have been performed for a slightly reduced strength
of the pairing force as compared to the estimate of~\cite{[Sat01a]}.
The  TRS and  $\delta$-force corrected  TRS
results are presented in fig.~\ref{fig05}.  The strength of the $\delta$-force
$g_{eff}\sim 650$\,MeV was estimated from a least square fit to
the data. The size of the matrix element is more or less consistent with
average liquid-drop estimate of $\sim 20/A$\,MeV of the
 $pn$ effect for valence particles in o-o nuclei although one
one would expect an enhancement in $N$=$Z$
nuclei purely due to geometrical reason (congruence effect).
However, the overall quality of the fit does not change very
much even if we double $g_{eff}$ and therefore is not
very conclusive.

The new elements in these calculations with respect to the one
presented in Ref.~\cite{[Sat01b]} can be summarized as follows:
({\it i\/}) the equilibrium deformation is calculated from the
TRS minimum  ({\it ii\/}) two quasiparticles of the same type
(i.e. not in time reversed states) have
been blocked and the residual
$pn$ interaction between the valence pair has been added
to the energy difference to reduce the coherence of
the ${\boldsymbol t}$=0 pair field.
These new calculations describe well the global decrease of $\Delta E(A)$
including the sign
inversion in the $f_{7/2}$ shell, see Fig~\ref{fig05}. However, the
details are far from being satisfactory reproduced. More extended studies of
the wave function of the valence nucleons is not expected
to improve the situation
since the TRS ground states are essentially spherical for light nuclei.
Deformation sets in only for heavy nuclei where spherical and
asymptotic matrix elements of the $\delta$-force are
anyhow very similar. This
seems to indicate that the $\boldsymbol t$=0 pairing
is too strong and needs further reduction.

\section{Influence of isovector particle-hole fields on the moment
of inertia in isospace}\label{isov}

  Phenomenological potentials [like the Woods-Saxon potential
  used here] depend only on the third component of the isospin
  $I=(N-Z)/A$. It means that essentially no modification of the
  mean-field is  done as a function of excitation energy in a given
  nucleus. On the contrary, such modifications are automatically
  included in self-consistent approaches through the changes
  of the isovector densities.

\begin{figure}
\begin{center}
\leavevmode
\epsfysize=10.0cm
\epsfbox{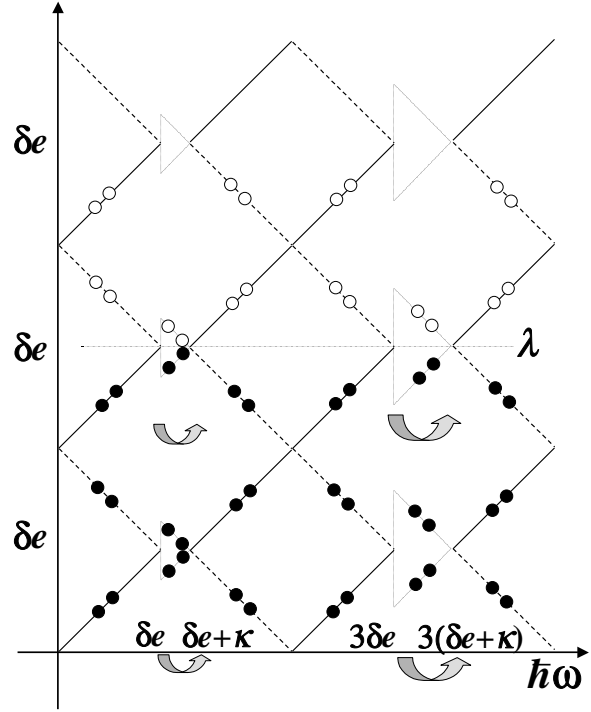}
\end{center}
\caption[]{
Single particle routhians representative for a model described
by the Hamiltonian (\ref{mfham}). Filled circles mark occupied
states. Arrows indicate shifts in crossing frequencies due
to the isospin dependence introduced via the
$\frac{1}{2}\kappa {{\hat{\boldsymbol t}}}\cdot {{\hat {\boldsymbol t}}}$
interaction.
}
\label{fig06}
\end{figure}

  The presence of a repulsive isovector mean-field provides
  an alternative [or additional] mechanism to lower the MoI in
  isospace. Let us illustrate it by using the simple iso-cranked
  single-particle model discussed in detail in
  Ref.~\cite{[Sat01a],[Sat01b],[Sat01c]}
  and including an additionally repulsive
  two-body interaction
  $\frac{1}{2}\kappa {{\hat{\boldsymbol t}}}\cdot
 {{\hat {\boldsymbol t}}}$
  analyzed by Neerg{\aa}rd~\cite{[Nee02]}:
\begin{equation}\label{nerh}
  \hat H^{\omega} = \hat h_{sp} - \vec{{\boldsymbol \omega}}
  \vec{{\hat {\boldsymbol t}}} + \frac{1}{2}\kappa {{\hat {\boldsymbol t}}}
  \cdot {{\hat {\boldsymbol t}}}
\end{equation}
  Linearization of the Hamiltonian (\ref{nerh})  and simoultanous assumption
  of one-dimensional rotation [say around $x$-axis, then
   $\langle \hat t_y\rangle =
   \langle \hat t_z\rangle =0$] leads effectively
  to a mean-field one-dimensional cranking Hamiltonian:
\begin{equation}\label{mfham}
  \hat H_{MF}^{\omega} =  h_{sp} -
  ( \omega - \kappa \langle {\hat t_x}\rangle ){\hat t_x}
\end{equation}
  with an effective isospin dependent cranking frequency.  The role
  of the isovector field is depicted schematically in fig.~\ref{fig06} where
  for simplicity the  equidistant single particle (s.p.) spectrum
  $e_i = i\delta e$ for $h_{sp} =\sum_{occ} e_i$ is assumed.
  It is clearly seen from the figure that isovector field simply shifts
  crossing frequencies from $\delta e, 3\delta e, 5\delta e, \ldots$ to
   $\delta e +\kappa , 3(\delta e+\kappa), 5(\delta e+\kappa), \ldots$
  These shifts are marked by dotted lines in fig.~\ref{fig06} since
  they are in fact instantenous due to non-collective
  iso-cranking. The  energy dependence $E(T)$ as a function of
isospin [$T=T_x\equiv\langle \hat t_x\rangle$] reads:
\begin{equation}\label{xxx}
         E(T) = \frac{1}{2}(\delta e + 2\kappa) T^2
\end{equation}
  i.e. is analogical to the one derived in~\cite{[Sat01a]} but with
  a reduced MoI-i.
  Adding standard  ${\boldsymbol t}=1, t_z=\pm 1$
  pairing with $\Delta_n = \Delta_p$ {\it deformes\/} the
  system [see Sec.~\ref{t1}]  and smoothes out the single-particle,
  step-like alignment process [{\it collective iso-rotation\/}]
  but does not essentially change the MoI-i as already discussed in
  \cite{[Sat01a],[Sat01b],[Sat01c]}
  [see also Sec.~\ref{t1pn}]. The {\it collectivity\/} effect
  introduced by  the ${\boldsymbol t}=1, t_z=\pm 1$ pairing
  correlations allows for incorporating isospin fluctuations
  in a static way through standard cranking condition
\begin{equation}\label{yyy}
         E = \frac{1}{2}(\delta e + 2\kappa) T^2 \longrightarrow
          \frac{1}{2}(\delta e + 2\kappa ) T(T+1)
\end{equation}
  and thus restores the linear [Wigner] term. This regime of the model
  with fixed $A$, and $T_z=0$ can be called the
{\it vertical} excitation regime.
  Changing the iso-cranking
generator from $\hat t_x$ to $\hat t_z$ and limiting
  ourselves to total particle number conservation $A$ only brings
  our model to the regime of {\it horizontal\/} excitations describing
  ground states of neighbouring nuclei of the same $A$. The mathematics
  used to solve both models is identical. Note, however, that the physics
  interpretation changes. For example, the physical restrictions for
  allowed s.p. [or quasiparticle] excitations which were related to
  iso-signature conservation and time reversal
  symmetries~\cite{[Sat01a],[Sat01b],[Sat01c]} now can be interpreted
  in terms of neutron and proton number parities. The cranking
  frequency measures the  difference between neutron and proton
  Fermi energies.

In the derivation of eqs.~(\ref{xxx}) the isovector Hartree
potential $V_T = \kappa \langle \hat{\boldsymbol t} \rangle
\hat{\boldsymbol t}$ was used rather than the two-body interaction.
This seems to be consistent with standard potential-like treatment of
the isovector terms in phenomenological nuclear potentials.
In such a case use of the cranking condition~(\ref{yyy}) also makes sense.
Since for large isospins $pn$ pairing is irrelevant, the inertia
parameter [the symmetry energy strength $a_{sym}$] is fully determined
by the mean-level density and $\kappa$.
The data does not give any signature of enhancement of
$a_{sym}$ in $N$=$Z$ nuclei. Still isoscalar
$pn$-pairing  effects may be related to the enhancement of
the {\it linear\/} term to $\frac{1}{2}a_{sym}T(T+x)$ with $x>1$.

The symmetry energy strength, and indirectly the value of $\kappa$,
can be conveniently fitted using the so-called double difference
$V_{pn} \approx \frac{\partial^2 B}{\partial N \partial Z}$
formula~\cite{[Zha89]}.
The results of a local fit to  $N\sim Z$ nuclei involving
$Z\geq 10$, $1\leq T_Z \leq 3$ nuclei [except o-o $T_Z$=1 nuclei]
are shown in table~\ref{fit}. This fit assumes
$E_{sym}\propto (N-Z)^2$ since the linear term vanishes because of
the second derivative. However, the strength of the linear  term
[or the Wigner energy $E_{wig}\propto |N-Z|$] can be fitted
separately using the prescription of Ref.~\cite{[Sat97a]}
which gives information about the local enhancement factor
$x$ in the $E_{sym}\sim T(T+x)$ formula.

\begin{table}
\begin{center}
\begin{tabular}{rrrrrc}
\hline
\hline
   $\alpha$  & $a_W$  &  $\sigma_{n-1}$  &  2$a_{sym}$ &  $\sigma_{n-1}$ &
   $x = a_W/2a_{sym}$ \\
\hline
\hline
0.95  &  39  & 0.196  &  31  &  0.106    &   1.26   \\
\hline
1/2  &    8  & 0.239  &   6  &  0.153    &   1.33   \\
2/3  &   14  & 0.213  &  11  &  0.125    &   1.27   \\
  1  &   47  & 0.196  &  38  &  0.107    &   1.24   \\
\hline
\hline
\end{tabular}
\caption[A]{
Results of the least-square fit for the Wigner energy strength
$E_{wig}=a_W|N-Z|/A^\alpha$ and for the symmetry energy strength
$E_{sym}=a_{sym}(N-Z)^2/A^\alpha$. The
minimum of the mean standard deviation is reached at $\alpha\sim 0.95$ for
both Wigner energy and symmetry energy fits. Results of one-dimensional
fits at fixed $\alpha = 1/2; 2/3; 1$ are given for comparison.}
\label{fit}
\end{center}
\end{table}

 The obtained strength of the symmetry energy lies in between
 large scale fits of $E_{sym}\sim T^2$ type~\cite{[Mol95]} and
 with those assuming $E_{sym}\sim T(T+1)$. The latest gives:
\begin{equation}\label{exp}
E_{sym} = \left( 134.4-\frac{203.6}{A^{1/3}} \right) \frac{T(T+1)}{A}\,\mbox{MeV}
\approx \frac{1}{2} 160  \frac{T(T+1)}{A}\,\mbox{MeV}
\end{equation}
 for  $A\sim 50$~\cite{[Duf95]}.  The data show clear enhancement of
 the linear term with $x\sim 1.25$ [last column] which is consistent with
 the early findings of J\"anecke~\cite{[Jan65]}.
 Indeed, it clearly leaves  room for isoscalar pairing since
 isospin fluctuations in the static theory limit gives
 $x=1$ by definition.

 We have demonstrated recently~\cite{[Sat02]} that the schematic
 interaction (\ref{nerh}) captures many features of realistic
 effective isovector interactions. In particular, in the
 Hartree-Fock limit, it gives rise to a {\it linear term\/}
 in the symmetry energy:
\begin{equation}\label{lta}
      \frac{1}{2}\kappa \langle \Delta {\hat {\boldsymbol t}}^2 \rangle
    \stackrel{HF}{\longrightarrow} \frac{1}{2}\kappa T.
\end{equation}
 This term is only due to the isovector part of mean HF potential
 and therefore represents only a fraction of
 the {\it linear term\/}. The
 mean-level density related contribution
\begin{equation}\label{ltb}
      \frac{1}{2}\varepsilon \langle \Delta
      {\hat{\boldsymbol t}}^2 \rangle
\end{equation}
seems to go beyond the mean-field approximation and its microscopic
evaluation requires RPA calculations~\cite{[Nee02]}.

Standard $pp/nn$ pairing leads to a strong quenching of the linear
term (\ref{lta}), as compared to its {\it sp\/}
estimate~\cite{[Sat02]}.
Since for a schematic force $\kappa$ is fixed
the suppression of this term is entirely
due to enhancement of the
$T/\langle \Delta {\hat {\boldsymbol t}}^2 \rangle$
ratio. This mechanism may also quench the linear term (\ref{ltb}).
Therefore, from microscopic point of view, one would expect $x<1$ i.e. below
the static estimate. For example,
Neerg{\aa}rd~\cite{[Nee02]} gives $x\sim 0.8$ for
$T=2$, $A=48$ nuclei. Let us point out, however,
that his estimate is based on a mean-level splitting
deduced from Fermi gas model which is unrealistically large
as shown in~\cite{[Sat02]}.

\begin{figure}
\begin{center}
\leavevmode
\epsfysize=8.0cm
\epsfbox{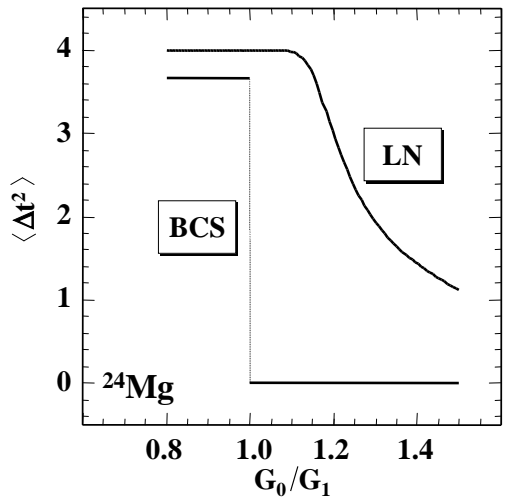}
\end{center}
\caption[]{
Dispersion in isospin $\langle\Delta{\hat {\boldsymbol t}}^2 \rangle$ calculated
for the LN and BCS models as a function of the ratio of the strengths
of isoscalar to isovector pairing forces. Note that $\boldsymbol t$=0 pairing
strongly quenches $\langle\Delta{\hat{\boldsymbol t}}^2 \rangle$. In particular,
in the limit of pure isoscalar pairing (BCS)
$\langle\Delta{\hat{\boldsymbol t}}^2 \rangle$=0.
}
\label{fig07}
\end{figure}

  Moreover, Neerg{\aa}rd's model does not
  take into account $\boldsymbol t$=0 pairing correlations. These
  correlations, even by entering at the dynamical RPA level, are
  expected to reduce isospin
  fluctuations as they in fact do in the static BCS or LN theory,
  see fig.~\ref{fig07}.
In conlusion, as expected, the
  presence of an isovector particle-hole field will
  definitely reduce  isoscalar pairing effects but it does not rule
  out their existence in $N\sim Z$ nuclei.

\section{Summary and conclusions}\label{sum}

We present analytical non-selfconsistent solutions
for a model including schematic isovector and isoscalar
pairing and discuss in detail the response of
the ${\boldsymbol t}$=1 and ${\boldsymbol t}$=0 pair-fields to
rotations in isospace in $N$=$Z$ nuclei. We are particularly
interested in the relations of the gauge angles of
the ${\boldsymbol t}$=1 pair-gaps and the position of
cranking axis in isospace. These relations decide upon the
character of rotation in isospace. In particular, it is shown
that within the standard model including $nn$ and $pp$
pair correlations and under the assumption of eq.(\ref{eq12})
no tilted solutions are possible.
Isorotation is either of collective type
$\vec {\boldsymbol \Delta} \perp \vec {\boldsymbol \omega} $
or non-collective type
$\vec {\boldsymbol \Delta} \parallel \vec {\boldsymbol \omega} $.
Since isorotation in $N$=$Z$ is planar, the $pn$
${\boldsymbol t}$=1 pair-field always induces collectivity.
Again it is shown that the most general solutions are obtained
for pure collective cases when the ${\boldsymbol t}$=1 pair-field
[${\boldsymbol \Delta}_\perp , {\boldsymbol \Delta}_{pn}$]
is perpendicular to the isocranking axis i.e. when
$\vec {\boldsymbol \Delta}_\perp \perp \vec {\boldsymbol \omega} $.
Other solutions might be possible only for particular
values of the gaps.

We also demonstrate that, within the simple pair-models,
mixing of ${\boldsymbol t}$=1 and ${\boldsymbol t}$=0
correlations is essentially forbidden or more precisely
restricted to a very special combination of pair gaps.
This is due to the different phase structure of eigenvectors 
among time-reversed states, compare eq.~(\ref{ev}) and 
eq.~(\ref{evt0}), or alternatively, due to 
the different transformation properties of 
${\boldsymbol t}$=1 and ${\boldsymbol t}$=0 gaps
under time-reversal.

Numerical calculations
show that this mixing is allowed within the HFB 
approximation but only for more elaborate
forces~\cite{[Ter98],[Goo99]}. It is also possible for schematic forces, 
within the LN approximation~\cite{[Sat97]}. In the latter
approximation, however, the $pn$ ${\boldsymbol t}$=1 pair gap vanishes 
which makes it unsuitable to
calculate the $T$=1 pair transfer between the $T$=1 o-o ground state and
$T$=0 e-e vacuum. Let us point out very clearly that, since 
$T$=1 o-o ground state and $T$=0 e-e vacuum are according
to our interpretation both 0QP states, 
the $\boldsymbol t$=1 pair transfer 
can in principle become enhanced through
${\boldsymbol t}$=1 $pn$ correlations as it is demonstrated within 
the BCS approximation in Sec.~\ref{tran}. 
On the other hand,  
the different fundamendal structure of the $T=0$ vaccua in e-e 
and o-o nuclei results in a quenching of the $T$=0 pair transfer 
even in the presence of strong $\boldsymbol t$=0 pairing.

The energy difference of $\Delta E_T \equiv E_{T=0} - E_{T=1}$ 
can be reproduced in a schematic manner by our extended TRS
calculations, including a residual $pn$-force of $\delta$-type.
Some discrepancies remain, most likely related to the fact that
the $\boldsymbol t$=0 pairing is too strong in our calculations
due to the lack of particle-hole isovector interaction. Inclusion
of such an  interaction will result in a weakening of the isoscalar
pairing. However, the enhancement in binding energy in $N$=$Z$
as compared to $N$-$Z$=2 nuclei seems to leave quite 
some room for $\boldsymbol t$=0 pair correlations as discussed
in Sec.~\ref{isov}. Further work along this line is in progress.

\bigskip

  This work was supported by the G\"oran Gustafsson Fundation, the Swedish
Institute, the Polish Committee for Scientific Research (KBN) under
 Contract No. 5~P03B~014~21, and  by the 
 U.S. Department of Energy under Contract Nos. DE-FG02-96ER40963
(University of Tennessee), and DE-AC05-00OR22725 (UT-Battelle).



\end{document}